\documentclass[fleqn,twoside]{article}
\usepackage{espcrc2}
\usepackage{alltt}

\title{Using the Mass Storage System at ZIB within I3HP}

\author{H.~St\"uben\address{Konrad-Zuse-Zentrum f\"ur
    Informationstechnik Berlin,
    Takustr.~7, 14195 Berlin, Germany}\thanks{Talk given at the Workshop
    on Computational Hadron Physics, Nicosia, Cyprus, 14--17 September 2005.}
and S.~Wollny\addressmark}

\begin{document}

\begin{abstract}
In the framework of I3HP there are two Transnational Access Activities
related to Computational Hadron Physics.  One of these activities is
access to the mass storage system at Konrad-Zuse-Zentrum f\"ur
Informationstechnik Berlin (ZIB).  European lattice physics
collaborations can apply for mass storage capacity in order to store and
share their configurations or other data (see \texttt{http://www.zib.de/i3hp/}).
In this paper formal and technical aspects of usage as well as the
conformance to the International Lattice DataGrid (ILDG) are explained.
\vspace{1pc}
\end{abstract}

\maketitle
\enlargethispage{\baselineskip}
\section{I3 HADRON PHYSICS}

The \emph{HadronPhysics Integrated Infrastructure Initiative (I3HP)}
is a project that originates from a joint initiative of over 2000
experimental, theoretical and computational physicists working in the
field of hadron physics \cite{i3hp}.  I3HP is funded by the European
Commission in the Sixth Framework Programme.  The project is structured
into nine \emph{Transnational Access Activities}, seven
\emph{Networking Activities}, and twelve \emph{Joint Research
Activities}.  There are three activities that are related to lattice
QCD: the Networking Activity \emph{Computational Hadron Physics},
Transnational Access to supercomputer resources at NIC (J\"ulich)
\cite{access-nic}, and Transnational Access to mass storage capacity
at ZIB (Berlin) \cite{access-zib}.  NIC is one of three national
German supercomputer centres, ZIB runs the supercomputer centre of the
federal state of Berlin.

\section{TRANSNATIONAL ACCESS}

The idea of Transnational Access is to give foreign researchers access
to important, typically experimental facilities.  It gives
experimentalists the opportunity to carry out interesting experiments
at facilities that they usually cannot use.  The access activities
related to computational QCD have a similar intention and provide
access to computational facilities.  While experimentalists have to
travel to the corresponding laboratories, computational facilities can
be accessed via the internet.

Researchers who want to use a facility have to write a scientific
proposal that is being peer reviewed.  How to apply for using the mass
storage system at ZIB is explained on our web page \cite{access-zib}.
It is not necessary to write an application for just downloading
configurations.  For downloading, a certificate is required (see section  
\ref{concepts}) and special software has to be used (see
section~\ref{software}).

\vspace*{-\baselineskip}
\section{LATTICE DATAGRIDS}

In large scale lattice QCD projects one typically stores gauge field
configurations and propagators.  Gauge fields are stored permanently
while propagators are stored for a limited period of time.
Propagators require much more storage space than configurations.
Hence they are kept at the site hosting the computer on which they were
calculated.  On the other hand, QCD gauge field configurations are
smaller and, in the case of dynamical fermions, much more expensive to
generate.  This has lead to the idea of sharing configurations in
order to fully exploit them.

The \emph{International Lattice DataGrid (ILDG)} \cite{ildg} was started
to make QCD gauge field configurations available at an international
level.  So far, ILDG infrastructures are being built up in Japan, UK,
USA, and Germany.  The German ILDG Grid is called \emph{LatFor DataGrid
(LDG)} \cite{ldg}.  ILDG develops standards for data formats and a
common middleware.  The definitions of a metadata format and a binary
file format are completed (see \cite{ildg}).  The \mbox{LatFor} DataGrid
conforms to the ILDG standards.

For this Transnational Access Activity we have decided to integrate
the storage system at ZIB into the broader LDG and ILDG activities.
Hence, the LatFor DataGrid became a joint effort of DESY (Hamburg and
Zeuthen), NIC (J\"ulich) and ZIB (Berlin).

\section{COMPONENTS OF LDG}

The main hardware components are \emph{storage elements (SE)} that
have \emph{hierarchical mass storages systems (HSM)} attached.  Each
site (in Berlin, Hamburg, J\"ulich, and Zeuthen) operates such a
storage element.  The hierarchical mass storages systems are large
tape libraries that work with tape robots.  The total capacity of the
HSM system at ZIB is about 1.2 PetaByte.  Data stored on the system is
very safe because there always exist two copies on different tapes.

The storage elements are small servers that run the dCache software
\cite{dcache} which was developed by DESY and Fermilab for storing
huge amounts of data distributed among heterogeneous server nodes.
From a user's perspective the distributed storage element servers
provide a single virtual filesystem tree.  Data may reside in the
server's disk cache or might be migrated to tape.  The dCache software
performs data exchanges to and from the attached tape libraries
automatically and invisibly to the user.

Besides the storage back-ends there are user interfaces at the
front-end.  In order to set up a Grid infrastructure, there are
several software components (middleware) needed in addition.  These
components are a \emph{virtual organisation}, \emph{Grid information
services}, a \emph{file catalogue}, an \emph{access control service},
and a \emph{metadata catalogue}.  At the middleware level LCG-2
software is used supplemented by developments of DESY.  LCG is the
\emph{Large Hadron Collider (LHC) Computing Grid} \cite{lcg}.

A virtual organisation (VO) is an organisational unit in a Grid
infrastructure. The VO representing the LatFor DataGrid is called
\emph{ildg}.  Grid information services handle e.g. authentication.
The file catalogue maps logical filenames to physical locations and
manages replicas of files.  The access control service (ACS) handles
access permissions. In LDG not all data is necessarily public. The
ACS allows to store public and private data in the same environment.
The metadata catalogue holds the metadata and makes it possible to
list metadata and perform search operations.

\section{IMPORTANT CONCEPTS}  \label{concepts}

Three important concepts for the usage of the hard- and software
infrastructure shall be explained: (1) authentication via certificates,
(2) logical filenames and physical locations of files, and (3) data
formats.

\subsection{Authentication}  \label{sect-authentication}

In a Grid context users are authenticated by presenting a certificate,
which is similar to the private/public key concept of the secure shell.
The exact technical procedure for obtaining a certificate depends on the
\emph{certificate authority (CA))}.  CAs for this Grid are listed in
\cite{lcg-certs}.

In any case, there are three basic steps that have to be done (cf.\ the
documentation of your CA).  First, one has to create a certificate
request file and a corresponding personal key.  The personal key should
be carefully protected with a secure passphrase and backed up.  Second,
the request file has to be sent to the responsible CA in a secure way.
In general personal authentication by presenting an identity card or
passport is required.  Third, one has to install the certificate that
one receives from the CA and the personal key on the machine where an
LDG user interface (UI) is installed.

With a valid certificate, one can use a so called \emph{grid-proxy},
which will do all necessary authentication in the background for a
given period of time.  So one does not have to type the passphrase every
time.  Usually, a certificate is valid for one year.  In general,
renewal is a much easier process than obtaining the initial one.

\subsection{Naming}

\newlength{\SpalteA}
\newlength{\SpalteB}
\settowidth{\SpalteA}{physical location}
\setlength{\SpalteB}{\textwidth}
\addtolength{\SpalteB}{-\SpalteA}
\addtolength{\SpalteB}{-4\tabcolsep}

\begin{table*}[t]
\caption{Examples of names.}
\label{tab-names}
\begin{tabular*}{\textwidth}{p{\SpalteA}l}
\hline 
object & name \\
\hline
ensemble & 
\verb+www.lqcd.org/ildg/qcdsf/nf2_clover/b5p29kp13500-16x32+ \\

configuration & 
\verb+/grid/ildg/qcdsf/nf2_clover/b5p29kp13500-16x32/qcdsf.515.00320.lime+ \\

physical location & 
\verb+srm://dcache.zib.de/pnfs/zib.de/data/ildg/\+\\
 & \verb+           qcdsf/nf2_clover/b5p29kp13500-16x32/qcdsf.515.00320.lime+ \\
\hline
\end{tabular*}
\end{table*}

\begin{table*}[t]
\caption{Array declarations corresponding to the storage sequence of 
SU(3) gauge fields. \texttt{Lx}, \texttt{Ly}, \texttt{Lz}, \texttt{Lt} are the
extensions of the lattice, \texttt{dim} = 4 is the dimension of space-time,
and \texttt{Ncolour} = 3 is the dimension of the SU(3) matrices.}
\label{tab-U}
\begin{tabular*}{\textwidth}{p{\SpalteA}l}
\hline
language & array declaration\\
\hline
C       & \verb|double U[Lt][Lz][Ly][Lx][dim][Ncolour][Ncolour][2];| \\
Fortran & \verb|complex U(Ncolour, Ncolour, dim, Lx, Ly, Lz, Lt)| \\
\hline
\end{tabular*}
\end{table*}

\begin{table*}[t]
\caption{Overview of \emph{ltool} commands.}
\label{tab-ltools}
\begin{tabular*}{\textwidth}{p{\SpalteA}p{\SpalteB}}
\hline
command            & description \\
\hline
\texttt{lget}      & Getting a binary of a configuration or the metadata
for an ensemble or a configuration from the Grid.
\\[\smallskipamount]
\texttt{lput} & Putting a binary and the corresponding metadata on the
Grid. It is not allowed to place a binary without corresponding
(syntactically-) correct metadata. Operation will take place in one
transaction, i.e., the data is either stored by successfully finishing
the operation or nothing will be stored if the operation fails
\\[\smallskipamount]
\texttt{lls} & Lists all configurations of an ensemble sorted by their
LFN (can also be used to show all ensembles in the MDC by using the
\texttt{--all} option).
\\[\smallskipamount]
\texttt{linit}     & 
Initialises a new ensemble in the MDC (requires administration rights).
\\[\smallskipamount]
\texttt{lupdate}   & 
Updates metadata in the MDC (has to be valid QCDml)
\\[\smallskipamount]
\texttt{lvalidate} & 
Check conformance of metadata to QCDml\\
\hline
\end{tabular*}
\end{table*}

Table~\ref{tab-names} shows examples of three types of names that are used in
the context of LDG.  This section explains these meaning of the names and
conventions for forming them.

When retrieving data (downloading gauge field configurations) from the
Grid one has to specify a \emph{logical filename (LFN)}.  The Grid
middleware translates the LFN into a physical location.  There may exist
multiple copies of the data in the Grid, so called replicas.  The
middleware is supposed to find the best available copy.

On uploading a configuration one has to specify an ID for the ensemble to
which the configuration belongs, which is called \emph{MarkovChainURI}, and a
\emph{physical location}.  The URI (unified resource identifier) is unique in
the world.  The physical location can be considered as an absolute path to the
configuration file within LDG.  The physical location is the place where the
data is actually stored.

Looking at the naming conventions adopted in LDG, one can see from the
examples shown in Table~\ref{tab-names} that all three types of names have a
large part in common.  All names contain the \emph{virtual organisation}
\texttt{ildg}. In the case of the MarkovChainURI \texttt{ildg} is strictly
speaking the last part of the URL \texttt{www.lqcd.org/ildg}. 
The uniqueness of the URL leads to the uniqueness of the MarkovChainURI.

After \texttt{ildg} follows the name of the collaboration which has
generated the
data, in the examples \texttt{qcdsf}.  The parts of the names that follow are
chosen by the collaboration.  However, the structure should be such that the
next part of the names represents a project and the part after that
represents an ensemble.  The name of a configuration and the physical location
have a file name in addition.

In the examples \verb+nf2_clover+ stands for the $N_f=2$ clover improvement
project.  The name of the ensemble \texttt{b5p29kp13500-16x32} repeats the
essential part of the metadata, i.e.\ $\beta = 5.29$, $\kappa = 0.13500$ on a
$16^3 \times 32$ lattice.  In the file name \texttt{qcdsf.515.00320.lime}
\texttt{515} is a job chain ID and \texttt{00320} is a trajectory counter.
As mentioned before, these three parts of the names were freely chosen by the
collaboration.

Syntactically the names are composed out of parts separated by slashes.
Basically, the physical location is a Unix file name specification,
i.e. a real directory structure is implied.  In the framework of ILDG it
would be allowed to chose names for ensemble, configuration, and
physical location completely independently.  For example, one could
think of using much shorter names for the LFN in order to facilitate
typing.  However, in LDG it was decided to essentially use the same
names at all levels for reasons of clarity.

\subsection{Data formats}

Data formats were defined by ILDG. There are conventions for formats
of metadata \cite{qcdml} and binary data \cite{qcdbin}. On uploading
binary data correct metadata have to be supplied.

\subsubsection{Metadata}

Within ILDG, a special metadata format, called QCDml, for the description
of configurations and ensembles has been defined \cite{qcdml}. QCDml
consists of two XML schemata, one for the description of an ensemble and
one for the description of a single configuration.  All metadata is
stored and exchanged in the XML format.  This allows a formal validation
before metadata is being uploaded to the \emph{metadata catalogue}.  For
example, a valid ensemble description has to have a
\verb|<markovChainURI>|-tag and must contain physical and algorithmic
information (see \cite{qcdml}). 

A valid configuration description must have a \verb|<dataLFN>|-tag,
which is the link to the binary file that can be downloaded from the
Grid.  In addition, a \verb|<markovChainURI>|-tag has to be provided,
which is the link back to the ensemble that it belongs to. Documentation
on how to markup configurations can be found in \cite{qcdml}.

\subsubsection{Binary data}

A binary file format for storing SU(3) gauge field configurations was
defined by ILDG.  This format is described in \cite{qcdbin}.  An ILDG
binary file consist of several parts that are packaged using the LIME
file format, which was developed by SciDAC.  A LIME API, utilities and
documentation are available from \cite{lime}.  With the API one can read
and write LIME files from a C programme.  Employing the utilities one
can pack or unpack LIME files at command line level.  One can also
extract individual files.

An SU(3) gauge field configuration packaged according to the conventions
of ILDG contains at least three files.  A file inside a package is
called a \emph{record}. These three files or record types, respectively,
are called:
\begin{verbatim}
   ildg-format
   ildg-binary-data
   ildg-data-LFN
\end{verbatim}
The record \texttt{ildg-binary-data} contains the gauge field
configuration. That record contains exactly the bytes of an array of floating
point numbers as it is declared in Table~\ref{tab-U}.  The floating point
format is IEEE and the byte ordering is big endian.   

The record \texttt{ildg-format} is an XML-document that contains the
precision of the floating point numbers (32 or 64 bit) and the lattice
size (see \cite{qcdbin} for the exact format).  Precision and lattice
size are also contained in the metadata.

The record \texttt{ildg-data-LFN} contains the logical file name as it
appears in the metadata.

A convenient way to extract records (files) from a LIME file is provided
by the utility
\begin{alltt}
 % lime_extract_type \emph{limeFile} \emph{recordType}
\end{alltt}
which writes the (first) record of the specified type to \emph{stdout}.

\section{USING LDG-SOFTWARE}  \label{software}

In the following sections, an overview of the LDG-software architecture
is given, followed by a short description of the user
interface (called \emph{ltools}) and a sample session.

\subsection{LDG-Software architecture}

Within the LDG community, there is a software bundle installable on any
computer running Linux. The bundle contains all the necessary software
to work with LDG (assuming you own a valid certificate). This includes
access to the \emph{meta data catalogue (MDC)} and all participating
\emph{storage-elements (SEs)}.  Actual access to the SEs is realised by
the use of the LCG-2 software \cite{lcg} and access to the MDC by the
use of a special client-software developed by DESY.  This complexity is
hidden from the user by offering a simple set of commands that combines
both (see the following section).  All software can be installed in a
simple way and without the need of root privileges.

Documentation and the software itself can be found in
\cite{ldg}. Initially, a packaging system called \texttt{lrpm} (which is
a kind of \texttt{rpm} tailored for the needs of LDG) has to be
installed and the location where you want to install the software has to
be defined. After this, installation, initialisation, updating or
deleting of the software is very easy, as only one or a few commands
have to be typed for each purpose.

\subsection{User tools}

The authors have written a set of easy to use command line tools called
\emph{ltools} which are distributed as part of the LDG-software package
\cite{ldg}. On the one hand, the motivation was to simplify the usage of
the corresponding LCG commands within the LDG context by using natural
defaults, combining sequences of commands, prevent the typical user from
erroneous use, providing better explanation- and error-messages and
making the software more configurable to personal needs. On the other
hand, the commands where designed to also access the MDC without the
need of an additional software for the user. In particular, the upload
of the data to the SE and the upload of the metadata to the MDC is
combined in a single transaction to circumvent inconsistencies in case
of an error.  An overview of the commands is given in Table~\ref{tab-ltools}.

\subsection{Sample session}

A typical session with \emph{ltools} looks as follows.  The
user starts by taking a look at which ensembles are stored in the MDC
by typing:
\begin{alltt}
   % lls --all
\end{alltt}
On the first call of an \emph{ltool} command \texttt{grid-proxy-init}
will be initiated automatically, so that the user is prompted for his
passphrase (unless the user has already a grid-proxy running).  The
output of \texttt{lls --all} is a list of MarkovChainURIs.  For each
MarkovChainURI one can get a list of LFNs of all configurations
for that URI by typing:
\begin{alltt}
   % lls \emph{MarkovChainURI}
\end{alltt}
One can download the metadata of a configuration by typing:
\begin{alltt}
   % lget -m \emph{LFN}
\end{alltt}
After inspection of the metadata, one might want to
download the actual configuration binary:
\begin{alltt}
   % lget \emph{LFN}
\end{alltt}
A user's guide with the full functional description of the commands can
also be found at \cite{ltools}.  Table~\ref{tab-session} shows the
output of an \texttt{lget} execution for downloading a configuration binary.

\begin{table*}[t]
\caption{Sample session. \texttt{grid-proxy-init} is automatically
  run at the first call of an \emph{ltool} command.}
\label{tab-session}
\begin{tabular*}{\textwidth}{p{\textwidth}}
\hline
\begin{alltt}
% \textbf{lget /grid/ildg/qcdsf/nf2_clover/b5p29kp13500-16x32/qcdsf.515.00320.lime}
Welcome to the Ltool-command  lget -
Testing grid-proxy-init:

Trying to start grid-proxy-init...
Your identity: /O=GermanGrid/OU=ZIB/CN=Hinnerk Stueben
Enter GRID pass phrase for this identity:
Creating proxy ................................................... Done
Your proxy is valid until: Wed Nov 16 03:56:22 2005
 
Trying to get binary ... 
Virtual Organisation is ildg 
Executing lcg-cp lfn:/grid/ildg/qcdsf/nf2_clover/b5p29kp13500-16x32/qcdsf.515.00320.l
ime --vo ildg file:/home/stueben/qcdsf.515.00320.lime 

Checking nonzero size of downloaded File ...ok.
\end{alltt}\\
\hline
\end{tabular*}
\end{table*}

\section{SUMMARY}
 
In this Transnational Access Activity a hard- and software
infrastructure was set up that is tailored to storing configurations
from simulations of QCD and that is well integrated into the
International Lattice DataGrid activities of the Computational Hadron
Physics community.

\section{ACKNOWLEDGEMENT}

We acknowledge the support from the European Community-Research
Infrastructure Activity under FP6 ``Structuring the European Research
Area'' programme (HadronPhysics, contract number RII3-CT-2004-506078).


\begin{thebibliography}{99}
\bibitem{i3hp}       \texttt{http://www.infn.it/eu/i3hp/}
\bibitem{access-nic} \texttt{http://www.fz-juelich.de/nic/i3hp-nic-ta/}
\bibitem{access-zib} \texttt{http://www.zib.de/i3hp/}
\bibitem{ildg}       \texttt{http://www.lqcd.org/ildg/}
\bibitem{ldg}        \texttt{http://www-zeuthen.desy.de/latfor/ldg/}
\bibitem{dcache}     \texttt{http://www.dcache.org/}
\bibitem{lcg}        \texttt{http://lcg.web.cern.ch/lcg/}
\bibitem{lcg-certs}  \texttt{http://lcg.web.cern.ch/LCG/users/registration/certificate.html}
\bibitem{qcdml}      \verb|http://www.ph.ed.ac.uk/ukqcd/community/the_grid/QCDml1.1/|
\bibitem{qcdbin}     \verb|http://www-zeuthen.desy.de/~pleiter/ildg/#filefmt|
\bibitem{ltools}     \texttt{http://www.zib.de/i3hp/ltools/}
\bibitem{lime}       \texttt{http://www.physics.utah.edu/~detar/scidac/}
\end{thebibliography}
\end{document}